\documentclass[aps,amsmath,amssymb,superscriptaddress,twocolumn,showpacs]{revtex4}
\usepackage{graphicx}
\usepackage{dcolumn}
\usepackage{bm}% bold math
\usepackage{colordvi}
\usepackage{color}
\usepackage{ulem}

\begin{document}

\title{Spin Chaos Manifestation in a Driven Quantum Billiard with Spin-Orbit Coupling}

\author{D.V. Khomitsky}
\email{khomitsky@phys.unn.ru}
\affiliation{Department of Physics, University of Nizhny Novgorod, 603950 Gagarin Avenue 23,
Nizhny Novgorod, Russian Federation}

\author{A.I. Malyshev}
\affiliation{Department of Physics, University of Nizhny Novgorod, 603950 Gagarin Avenue 23,
Nizhny Novgorod, Russian Federation}

\author{E.Ya. Sherman}
\affiliation{Department of Physical Chemistry, The University of the Basque
Country, 48080 Bilbao, Spain} \affiliation{IKERBASQUE Basque Foundation for
Science, Bilbao, Spain}

\author{M. Di Ventra}
\affiliation{Department of Physics, University of California, San Diego,
9500 Gilman Drive, La Jolla, CA 92093-0319, USA}

\begin{abstract}
The coupling of orbital and spin degrees of freedom is the source of many interesting phenomena.
Here, we study the electron dynamics in a quantum billiard driven by
a periodic electric field--a mesoscopic rectangular quantum dot-- with spin-orbit coupling.
We find that the spatial and temporal profiles of the observables
demonstrate the transition to chaotic dynamics with qualitative modifications of the power spectra
and patterns of probability and spin density.
The time dependence of the wavefunctions and spin density
indicates spin-charge separation seen in the decay of the spin-charge density correlators.
Experimental verification of this spin chaos effect can lead to
a better understanding of the interplay between spin and spatial
degrees of freedom in mesoscopic systems.

\pacs{72.25.Dc,72.25.Pn,73.63.Kv,75.70.Tj}

\end{abstract}

\maketitle

\section{Introduction}

The emergence of stochasticity is of fundamental importance for classical and quantum physics
with broad interdisciplinary connections and applications \cite{Gutzwiller,Reichl,Haake,Stockmann,Casati}.
Fascinating examples of irregular dynamics can be found in meso- and nanoscale
systems, including quantum dots \cite{Stockmann,Nakamura,Bushong}.
While the understanding of charge transport in such systems
is already quite deep \cite{book}, the knowledge of the chaotic spin evolution is still poor.
Since the related branch of physics, ``spintronics'' is among the most
promising research fields \cite{Awschalom,Dyakonov,Zutic},
it is of importance and applied interest to study the spin dynamics in mesoscopic
systems with coupled charge and spin degrees of freedom. A natural example for these studies is provided
by semiconductor structures where the spin-orbit coupling (SOC) plays a significant role in
dynamics, including the ability of spin manipulation
by electric field \cite{Rashba1,Nowack,Pioro,Golovach06,Levitov03}.
It has been predicted that for a sufficiently strong driving field the dynamics of charge and spin can become
unexpectedly complicated for electron in a double quantum dot \cite{KS2009,KGS2012,Chototlishvili}.
However, it is not known what kind of irregular coupled dynamics one may expect in a mesoscopic
structure such as a semiclassical quantum billiard.

Without SOC such structures may demonstrate certain classical traits
of the transition to chaos due to the high density of states,
including the formation of irregular wavefunctions inside the billiard \cite{Stockmann,Nakamura,BerggrenSadreev,BerggrenSadreev2}.

It is known that the eigenstates distributions in rectangular
billiards with SOC demonstrate fingerprints of chaos such as
the Wigner statistics, not expected for integrable systems \cite{Berggren}.
Since chaos driven by an external periodic field is a common phenomenon in nonlinear systems,
it is natural to ask whether an irregular motion arises in such a billiard
for {\it coupled} charge and spin channels under a periodic driving.
The problem itself is complicated since there are no explicit criteria for identifying
the chaotic regimes in the dynamics of quantum spin observables
not having classical counterparts.

In this paper, we present a model for exploring dynamical regimes of coupled charge and spin degrees of freedom for
two-dimensional (2D) electrons confined in a rectangular quantum billiard with Rashba SOC and
an in-plane magnetic field. We consider charge and spin dynamics driven by a periodic electric
field resonating with transitions between two nearest semiclassical size quantization levels.
We observe strong indications of transition to irregular dynamics both for spatial coordinates and
spins as studied using the Floquet stroboscopic technique \cite{Reichl,Stockmann,KGS2012,DM2002}.
Although the quantum nature of our systems hampers pure classical
manifestation of chaos like the Lyapunov exponents,
the scenarios for both charge and spin
evolution suggest the transition to irregular regimes. We found that the most sensitive observables
are the densities of charge and spin, where the textures of different
shapes determined by the driving frequency and amplitude are formed.
The results can be useful for the understanding of the quantum chaos involving spin
as well as for the design of semiconductor-based spintronics devices.

\section{Model}

Our Hamiltonian, $H(t)=H_0+V(t)$, consists of the unperturbed part $H_0$ describing an electron
confined in the 2D rectangular billiard
with sides $a$ and $b$, hard-wall boundary conditions, Rashba SOC and Zeeman interaction
to the in-plane magnetic field $B_x$:

\begin{equation}
H_0=\frac{p^2}{2m}+H_{\rm so}+\frac{1}{2} g\mu_B \sigma_x B_x,
\label{H0}
\end{equation}

where we take $H_{\rm so}=\alpha_{R}\left(\sigma_x k_y - \sigma_y k_x  \right)$ in the Rashba form.
Here $m$, $\alpha_{R},$ and $g$ are the electron effective mass, Rashba SOC constant,
and $g$-factor, respectively. The in-plane magnetic field lifts the Kramers degeneracy,
however, does not cause the diamagnetic coupling.
Since the $\sigma_x$ operator is present in both the SOC and Zeeman terms,
the Zeeman splitting is coupled to the orbital quantization making it orbital state-dependent.
The spin-coordinate entangled wavefunction of the $n$-th state of $H_{0}$
can be constructed as a superposition of
the orbital wavefunctions multiplied by the two-component spinors,

\begin{equation}
{\bm \psi}_n(\mathbf{r})=\sum_{l_{x},l_{y}}
\left[
\begin{array}{c}
\zeta^{\uparrow}_{l_{x},l_{y}}
\\
\zeta^{\downarrow}_{l_{x},l_{y}}
\end{array}
\right]_{n}
\cdot
\frac{2}{\sqrt{ab}} \sin \frac{\pi l_{x}x}{a} \sin \frac{\pi l_{y}y}{b},
\label{psin}
\end{equation}

where the $n$-dependent coefficients $\zeta^{\uparrow}_{l_{x},l_{y}}$
and $\zeta^{\downarrow}_{l_{x},l_{y}}$ are determined from the matrix eigenvalue problem,
and $\mathbf{r}=(x,y)$. The SOC leads to the spin-coordinate entanglement
of the state ${\bm \psi}_{n}(\mathbf{r})$, while at $\alpha_{R}=0$ the corresponding state is the
product of the orbital state and eigenstate of $\sigma_{x}$.

The typical level splitting in a $\mu$m-size billiard, being significantly lower than in a nanoscale
quantum dot, leads to a better possibility to reproduce the quantum-classical
correspondence in the regular and stochastic evolution. Berggren and Ouchterlony proved
that the SOC leads to the non-Poissonian level statistics, indicating
the presence of quantum chaos \cite{Berggren}.
However, this well-known picture of static eigenstate description
leaves open the question on the dynamical characteristics of the associated evolution.
In particular, on the difference between simply irregular and chaotic
behavior for coupled spin and charge
degrees of freedom, which we consider in our paper.
The external driving is needed since it allows manipulating the perturbation amplitude in a controlled manner,
which is required in most of the applications to transfer and keep the system in a required state
in the presence of inevitable momentum and spin relaxation.
We will see that the quantum-classical correspondence with the generation of the chaotic
behavior of observables will be maintained mainly in the initial period of the evolution, and after that more regular quantum dynamics
is achieved both for charge and spin degrees of freedom. This is in agreement with general properties
of quantum chaos \cite{Gutzwiller,Reichl,Haake,Stockmann,Casati}.

The driving term $V(t)=e {\cal E}_{0} x \cos \omega t$, where $e$ is the fundamental charge,
is chosen as the monochromatic uniform electric field of amplitude
${\cal E}_{0}$ which induces the local resonance between the level pair split by $E_{n_{0}}-E_{n_{0}-1}=\hbar \omega_0$ for a given $n_{0}$.
Due to the spin-coordinate entanglement in Eq. \eqref{psin}, the electric field causes transitions between states with different expectation
values of spin, and leads to nontrivial dynamics as discussed below.

To characterize the couplings in the system, we apply a perturbative approach and
find the dimensionless parameters for spin-orbit coupling and electric field strength.
We begin with the unperturbed spectrum:

\begin{equation}
E\left( l_{x},l_{y},\sigma_{x}\right)=\frac{\pi ^{2}\hbar ^{2}}{2m}
\left(\frac{l_{x}^{2}}{a^{2}}+\frac{l_{y}^{2}}{b^{2}}\right)+\frac{1}{2} g\mu_B \sigma_x B_x
\end{equation}

and take the main semiclassical term (assuming $l_{x}\gg1,l_{y}\gg1$) in the energy difference of the same-spin states

\begin{equation}
\Delta E\equiv E\left(l_{x}+1,l_{y},\sigma_{x}\right) -E\left(l_{x},l_{y},\sigma_{x}\right)
=\frac{\pi ^{2}\hbar ^{2}}{ma^{2}}l_{x}.
\end{equation}

For spin-orbit coupling we take the ratio

\begin{equation}
f_{\rm so}\equiv\frac{|\langle l_{x}+1,l_{y},1|H_{\rm so}|l_{x},l_{y},-1\rangle|}{\Delta E}=
\frac{2}{\pi^{2}}\frac{\alpha_{R}}{a}\frac{ma^{2}}{\hbar^{2}},
\label{fso}
\end{equation}
{ that is, essentially, the ratio of the billiard size to the spin precession length $\hbar^{2}/m\alpha_{R}$}
as a dimensionless strength of spin-orbit coupling. For the driving field we proceed similarly and define

\begin{equation}
f_{\cal E}\equiv\frac{e{\cal E}_{0}|\langle l_{x}+1,l_{y},1|x|l_{x},l_{y},1\rangle|}{\Delta E l_{x}^{-1}}=
\frac{2}{\pi^{4}}e\mathcal{E}_{0}a\frac{ma^{2}}{\hbar ^{2}}.
\end{equation}

\begin{figure}[tbp]
\centering
\includegraphics*[width=85mm]{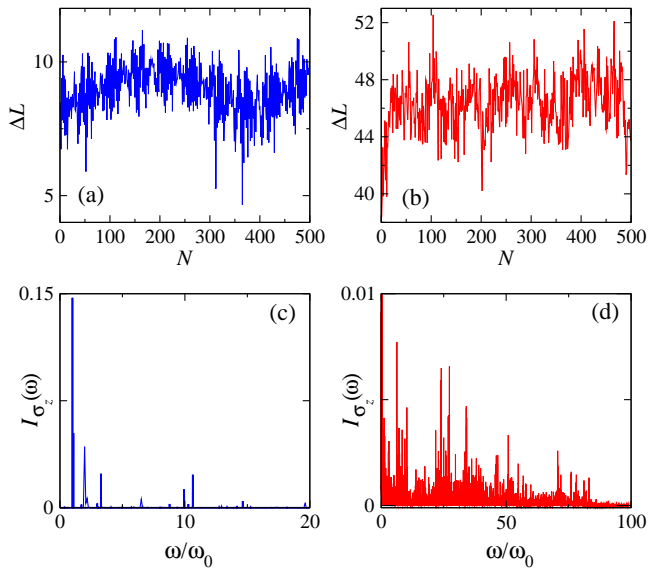}
\caption{(Color online) (a),(b) Stroboscopic evolution of the variance $\Delta L$ of the level number
effectively involved into the dynamics for $N=500$ periods,
$T$, of driving field, (a) weak driving electric field with amplitude ${\cal E}_{0}=0.14$ V/cm and
(b) moderate driving field ${\cal E}_{0}=0.70$ V/cm. After a short period of diffusive growth the evolution of $\Delta L$
approaches the stationary regime with stable average value.
(c),(d): Fourier power spectra (in dimensionless units) for the mean values of $\sigma_z$ spin component,
(c) ${\cal E}_{0}=0.14$ V/cm and (d) ${\cal E}_{0}=0.70$ V/cm.
The total observation time in (c) and (d) is $T_{\text{tot}}=100 T$.}
\label{figpowerspectra}
\end{figure}

\section {Level variance evolution and Fourier power spectra}

We begin by solving the nonstationary
Schr{\"o}dinger equation with the driving $V(t)$ in the basis of the eigenstates
(\ref{psin}) to obtain the spinor wavefunction

\begin{equation}
{\bm {\bm \Psi}}(\mathbf{r},t)=\sum_n C_n(t) {\bm \psi}_n(\mathbf{r}),
\label{wf}
\end{equation}

where the time-dependent coefficients $C_n(t)$ are solutions of a system of ordinary differential equations.
The initial condition is taken as the single level occupancy,
$C_n(0)=\delta_{nn_0}$. { The initial single level state can be prepared in a mesoscopic billiard
by resonant tunneling of an electron with required energy,
entering from attached leads \cite{restunn}.
If the initial state is a superposition of the eigenstates, its dynamics can be found as the corresponding
superposition of the time-dependent states demonstrated below.}

For numerical calculations we consider a GaAs billiard with (if not stated otherwise)
$\alpha_{R}=5$ meVnm, $a=2.1$ $\mu$m and $b=1.5$ $\mu$m (same as in Ref.\cite{Berggren}).
Here $\hbar^{2}/ma^2=0.26$ $\mu$eV, and $f_{\rm so}\approx 2.0$. We assume a magnetic field $B_{x}=500$ Gs
with the Zeeman splitting of 1.3 $\mu$eV. The typical driving frequency here is $\nu=\omega_0/2\pi=0.78$ GHz,
and the initial state is on the level $n_{0}=200$.

We calculate the evolution of quantum observables using the wavefunction (\ref{wf}).
The parameters describing the evolution in the Hilbert space can be chosen as
the mean level number ${\bar L}(t)=\sum_{L} L|C_{L}(t)|^{2}$ and its dispersion
$\langle\Delta^{2}L(t)\rangle=\sum_{L}\left( L -{\bar L}(t) \right)^2|C_{L}(t)|^{2}$. It is known that during the initial stages
of the development of quantum chaos the variance of the level number $\Delta L (t)\equiv\langle\Delta^{2} L(t)\rangle^{1/2}$
grows with time \cite{Reichl,DM2002}. This corresponds to the diffusion in the Hilbert space with increasing number of levels being involved,
and represents counterpart of the classical chaotic dynamics.
After some time, this stage transforms into the stabilized quantum evolution,
where $\Delta L (t)$ saturates, thus showing the suppression
of the Hilbert space diffusion \cite{Reichl,Casati,DM2002}.

The stroboscopic dynamics of level variance $\Delta L (t)$ is shown in Fig.\ref{figpowerspectra} for $500$ periods $T$, of driving
(a) for the moderate field ${\cal E}_{0}=0.14$ V/cm ($f_{\cal E}\approx2.3$) and (b)
for stronger field ${\cal E}_{0}=0.70$ V/cm ($f_{\cal E}\approx11.5$).
It is clear that the initial fast growth in the number of involved levels ceases after
$N=10\ldots 30$ periods of driving field, and after that $\Delta L(t)$ demonstrates the oscillating behavior around
the average $\Delta L_{\rm av} \approx 9$ in Fig. \ref{figpowerspectra}(a) and $\Delta L_{\rm av} \approx 46$
in Fig. \ref{figpowerspectra}(b).
This behavior corresponds to the expected growth in the number of participating states with increasing driving amplitude.
We may conclude that the dynamics rather quickly reaches a
stabilized regime without further diffusion in the Hilbert space, indicating that even a large
billiard with hundreds of levels involved in the dynamics behaves
essentially as a quantum system without the long-lasting Hilbert space diffusion.

Quantum evolution can be described in terms of the power spectrum,
defined for an observable $\xi(t)$ as

\begin{equation}
I_{\xi}(\omega)=\left| \int_{-\infty}^{+\infty} \xi(t) e^{-i \omega t} dt \right|^2.
\label{powerspec}
\end{equation}

Figures \ref{figpowerspectra}(c),(d) show power spectra for the spin component $\sigma_z$
for the same driving fields, plotted in dimensionless units for mutual comparison.
While for Fig. \ref{figpowerspectra}(c) the spectrum has the form of discrete harmonics, the stronger driving
field leads to qualitatively richer spectrum as shown in Fig. \ref{figpowerspectra}(d) where the lower band is filled
continuously. According to the basic concepts of quantum evolution,\cite{Gutzwiller} this may be considered as the onset of chaos.
{ It should be mentioned that the long-period patterns in level variations visible in Fig.\ref{figpowerspectra}(a),(b)
provide sizable contributions into the Fourier power spectra for the spin component in Fig.\ref{figpowerspectra}(c),(d)
as the strong peaks at the left part of the frequency axis near $\omega=\omega_0$.
However, the main difference between the regular and chaotic dynamics is in the mid- and high-frequency
part of the spectrum, as it can be seen by comparing Figs.\ref{figpowerspectra}(c) and (d),
where a densely filled frequency band in Fig.\ref{figpowerspectra}(d) represents the chaotic
behavior.}

\begin{figure}[tbp]
\centering
\includegraphics*[width=80mm]{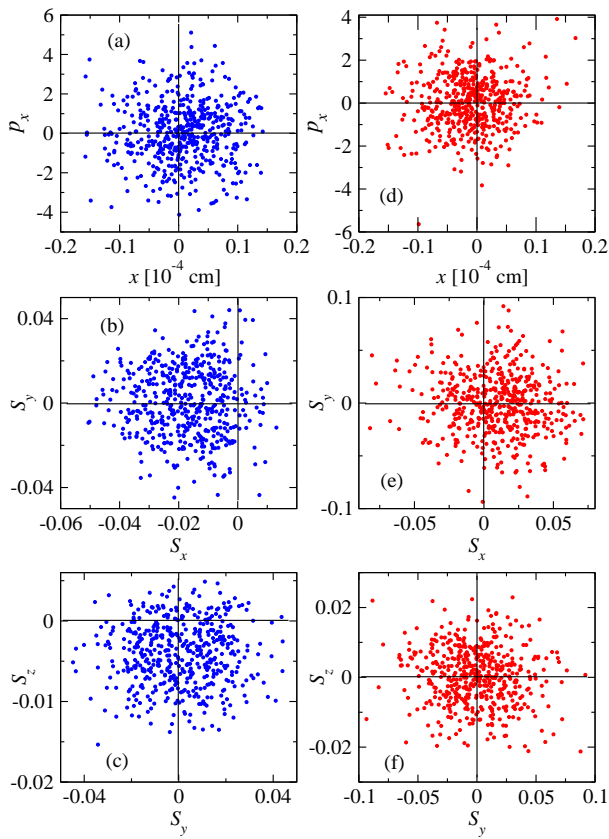}
\caption{(Color online) stroboscopic Poincar\'e plots: (a,d) coordinate mean values $(x(N), p_x(N))$ and (b) and (e) spin
mean values $(S_x(N),S_y(N))$ and (c) and (f) - spin mean values $(S_y(N),S_z(N))$, shown for
the driving field strength ((a)-(c)) ${\cal E}_{0}=0.14$ V/cm  and ((d)-(f)) ${\cal E}_{0}=0.70$ V/cm.
The coordinate degree of freedom evolves in the same manner as
the spin degree of freedom due to Rashba SOC, both indicating an irregular regime of dynamics.
Momentum $p_x$ is measured in units of $\hbar/\mu{\rm m}$.}
\label{PB}
\end{figure}

\section{Poincar\'e sections}

In order to gain insight into mutual impact of orbital and spin motion,
we plot the evolution of expectation values in the pair of canonical variables $(x(t), p_x(t))$ and in the
non-canonical pairs $(S_{\beta}(t),S_{\gamma}(t))$, where $\beta$ and $\gamma$ are Cartesian coordinates.
For a periodic driving it is of interest to consider
the Poincar\'e sections at stroboscopic times $t=NT$ (with $N$ an integer) such as $(x(N), p_x(N))$
or $(S_{\beta}(N),S_{\gamma}(N))$. To obtain the evolution with high accuracy, we use the Floquet
stroboscopic technique which requires the direct integration of the time-dependent
Schr{\"o}dinger equation only at a single period $T$ of the perturbation $V(t)$. After that
the state of the system at any $t=NT$ can be obtained by a finite algebraic procedure \cite{KGS2012,DM2002}.
In Fig. \ref{PB} we show the stroboscopic Poincar\'e plots for the same fields
as in Fig.\ref{figpowerspectra}.

Due to the Rashba SOC the coordinate degree of freedom evolves in the same manner as
the spin one, indicating an irregular regime of dynamics like a
``stochastic sea"\cite{Gutzwiller,Reichl} without any ``islands of regularity" with periodic orbits.
This example of dynamics may serve as a tool for observing the spin chaos in
quantum systems. In the absence of SOC, the states would remain the eigenstates
of $\sigma_{x}$, and the Poincar\'e sections would be reduced to the { points $S_x(N)=\pm1$ and $S_y(N)=S_z(N)=0$.}
Small expectation values of spin components are due to the spin-coordinate entanglement of the
states in Eq. \eqref{psin}.
It can be seen that the spin dynamics is strongly sensitive to the number of levels involved
each having a different spin polarization, so the amplitudes for the mean values also grow at higher electric fields.
By looking at Fig. \ref{PB} one also notices that the dynamics in coordinate space is not so sensitive to the driving
field strength as the evolution of spin variables. This can be explained by taking into account the structure of eigenstates
and the initial condition of our evolution model which is an eigenstate with a rather high number of spatial harmonics, and is completely
delocalized in the billiard.
The volume spanned by the evolution of the $(x,p_x)$ pair of mean values does not expand greatly with increasing driving strength since
these variables have comparable expectation values for two different numbers of levels effectively
involved into the dynamics for weak (Fig. \ref{figpowerspectra}(a)) and moderate (Fig. \ref{figpowerspectra}(b))
driving amplitudes.

It should be mentioned also that the average value of spin for the weaker driving
is shifted from zero more visibly, as it can be seen in Fig. \ref{PB}. We explain this
behavior by distinct, but generally alternating spin polarizations of
the basis states, where the greater number of states involved at stronger driving
leads to more effective canceling of non-zero contributions to the midpoint of spin
density stroboscopic ensemble.

\section{Spin textures and spin-charge separation}

A typical experiment with scanning of the billiard with electron gas
measures local spin density integrated over the ``spot" under the probe \cite{Ritchie}.
Thus, the evolution of local spin density in the billiard can be of interest for further experimental advances
in exploring and controlling various regimes of the driven spin dynamics.
We then look at the spatial distributions of spin density in the whole billiard, or spin textures, considered at
the stroboscopic time $t=NT$ with arbitrary $N$, together with the charge density contained in the spinor components.

\begin{figure}[tbp]
\centering
\includegraphics*[width=85mm]{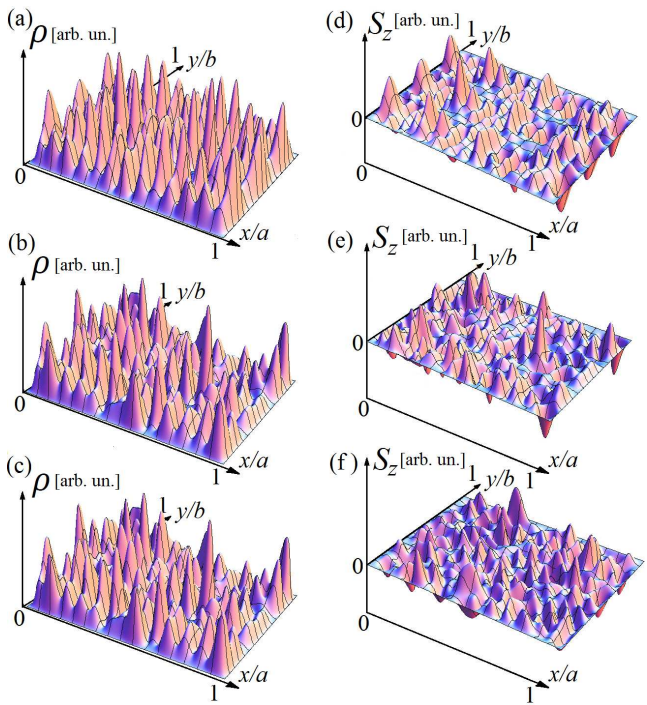}
\caption{ (Color online) Probability density distributions for
(a)-(c) the charge and (d)-(f) the $S_z$ component of spin density in the billiard
(a),(d) for the initial state taken as an eigenstate of the billiard with high number of spatial harmonics, and
 after $NT=500$ periods of driving field with amplitude (b),(e) ${\cal E}_{0}=0.14$ V/cm, and
(c),(f) ${\cal E}_{0}=0.70$ V/cm.
The spatial distributions for the $S_z$ (and other spin components) have the regular component on the medium
scale and the irregular peaked contribution on small scale for both initial state (a),(d) and after
the driven evolution (b,c,e,f).}
\label{figspindens}
\end{figure}

The charge- $\rho(\mathbf{r},N)$ and spin density components $S_{\beta}(\mathbf{r},N)$
are found with the wavefunction (\ref{wf}) as
\begin{eqnarray}
&&\rho(\mathbf{r},N)={\bm {\bm \Psi}}^{\dagger}(\mathbf{r},N) \, {\bm {\bm \Psi}}(\mathbf{r},N),
\label{rho}\\
&&S_{\beta}(\mathbf{r},N)={\bm {\bm \Psi}}^{\dagger}(\mathbf{r},N) \, \sigma_{\beta} \, {\bm {\bm \Psi}}(\mathbf{r},N),
\label{Sbeta}
\end{eqnarray}
respectively.
In Fig.\ref{figspindens} we show the probability distributions
for the charge and the $S_z$ spin component for the initial
eigenstate $(n=200)$ with high number of spatial harmonics, and after $N=500$ periods of driving.
The other components of spin density as well as the probability density have similar patterns.
The charge and spin densities calculated at $N<500$ indicate that the picture is stabilized after several hundreds
of driving periods, similar to the mean level dynamics in Fig.\ref{figpowerspectra}(a),(b).
We have found that the distributions presented in Figs.\ref{figspindens}(b),(c),(e),(f) are formed in the evolution
in the electric field as the interplay of two distinct patterns, which modify the regular structure of the initial state
in Fig.\ref{figspindens}(a),(d).
One pattern is the average-scale and large-scale structure with regular spatial oscillations stemming from the limited number
of basis states effectively involved into dynamics.
The other pattern of spin density in Fig. \ref{figspindens} has an irregular and spatially chaotic character,
mostly on the small and medium scales with formation of peaks with variable height.
The amplitude of small scale irregular contribution to the spin density grows with increasing driving strength.
This is an indication of the chaotic regime which is induced in our system at strong driving field.
We have observed the formation of similar two-scale density distributions for both charge and spin
also at longer times compared to the snapshot in Fig.\ref{figspindens}. The formation of this stable picture can
be attributed to the onset of the quasi-stationary profile of the dynamic regime for our driven evolution
after the period of initial quasiclassical chaotic-like regime. This can be observed, for example, in the Hilbert space
dynamics of level number shown in Fig.\ref{figpowerspectra}(a),(b). After the starting period of diffusion in the Hilbert space
described by the growing number of levels involved into dynamics the evolution is transformed into the quasi-periodic pattern with
the stable behavior of all the observables at arbitrary long times, including the spin textures shown in Fig.\ref{figspindens}.
Thus, we believe that these stable and predictable two-scale density patterns with both regular and irregular contributions
can be observed in scanning probe experiments performed in various time frames.

In fact, such complicated spatial profiles of probability density are known in quantum systems like
billiards or waveguides demonstrating chaotic behavior \cite{Stockmann,Nakamura,BerggrenSadreev,BerggrenSadreev2,BS2004,Ngo2011,Urbina2013}.
Finally, from Fig. \ref{figspindens} one may notice a developing with the time difference
between the distribution of spin and charge, possibly indicating the spin-charge separation in this system.
Since the billiard is a simple rectangular,
we do not see any specific  ``scars" in the density usually
occurring in the chaotic billiards where chaos appears due to their shape \cite{BerggrenSadreev,BerggrenSadreev2,Ngo2011,restunn}.
In our billiard the chaos is generated by the SOC rather
than by the geometry \cite{Berggren}. After turning on the driving this initially chaotic structure
of the spectrum and the eigenstates determines the dynamical variables and density distributions such as the spin textures
in Fig.\ref{figspindens}.

\begin{figure}[tbp]
\centering
\includegraphics*[width=70mm]{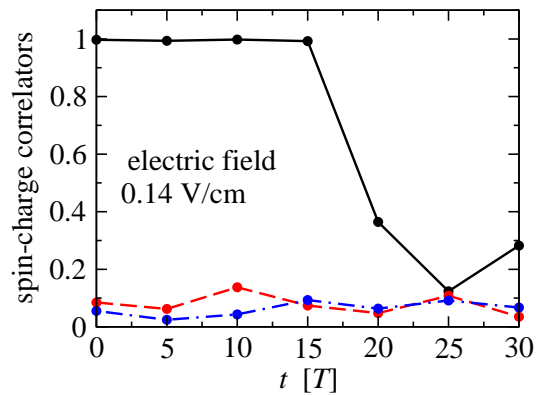}
\caption{(Color online) The magnitude of the time-dependent correlator between
the charge density and the $S_{x}$ component of spin density for ${\cal E}_{0}=0.14$ V/cm;
$\alpha_R=0.5$ meVnm (solid line), $\alpha_R=1.5$ meVnm (dashed line), and $\alpha_R=5.0$ meVnm (dashed-dotted line).
The correlators fall below $0.2$ when SOC strength (\ref{fso}) is high, and the spin-charge separation occurs.}
\label{figcorrel}
\end{figure}

One of the tools for checking the degree of spin-charge separation is the correlators
between the charge- and the spin density components taken on
the spatial grid with ${\cal N}$ points and treated as statistical variables.
We then calculate the correlators
of quantities $q$ and $v$ as $r_{q,v}=K_{q,v}/w_{q}w_{v}$,
where
\begin{equation}
K_{q,v}=\sum_i \frac{q_i v_i}{\cal N}-m_{q} m_{v}
\end{equation}
is the correlation coefficient, $m_{q,v}$ is the mean value, and
$w_{q,v}$ is the corresponding variance.
An example of such a time-dependence is shown in Fig.\ref{figcorrel}.
The amplitudes of correlators are below $0.2$ when SOC is strong (\ref{fso}) mixing
several energy levels, and the spin-charge separation occurs.
This result can be viewed as another consequence of the strong SOC and driving producing the
time-dependent entanglement of the charge and spin degrees of freedom.

\section{Conclusions}

We have studied the electron dynamics in
a quantum billiard with spin-orbit coupling and driven by a monochromatic electric field.
It was found that the spatial and time resolved patterns for probability and spin densities
demonstrate the onset of chaotic dynamics with qualitative modifications of the power spectra
and spatial patterns. In particular, we have identified new regimes of quantum chaos
in this system described by two-scale spatial charge and spin density distributions.
The onset of spin-charge separation effect is predicted by the dynamics of the spin and charge density
correlators. Our predictions can be important for the understanding of the coupled spin-charge
transport through mesoscopic billiards driven by a uniform electric field, where chaos can arise
for both spin and charge current observables. The stationary density distributions seen in
the absence of the driving can be verified in the tunneling experiments similar to those presented in Ref.[\onlinecite{Ritchie}].

\section*{Acknowledgements}

The authors are grateful to A.F. Sadreev for stimulating discussions.
D.V.K. and A.I.M. are supported by the RFBR Grants No. 13-02-00717a, 13-02-00784a.
E.S. is supported by of the University of Basque Country UPV/EHU under program UFI 11/55, Spanish
MEC Grant No. FIS2012-36673-C03-01, and ``Grupos Consolidados UPV/EHU
del Gobierno Vasco" grant IT-472-10. M.D. acknowledges support from DOE grant DE-FG02-05ER46204.


\begin{thebibliography}{99}

\bibitem{Gutzwiller}
M.C. Gutzwiller, {\it Chaos in Classical and Quantum Mechanics},  Springer-Verlag, New York, 1990.

\bibitem{Reichl}
L.E. Reichl, {\it The Transition to Chaos. Conservative Classical Systems and Quantum Manifestations}, 2nd Ed.,
Springer-Verlag, New York, 2004.

\bibitem{Haake}
F. Haake, {\it Quantum Signatures of Chaos}, Springer-Verlag Berlin Heidelberg, 3rd Ed., 2010.

\bibitem{Stockmann}
H-J. St{\" o}ckmann, {\it Quantum Chaos: An Introduction}, Cambridge University Press, 1999.

\bibitem{Casati}
G. Casati, B.V. Chirikov, F.M. Izrailev, and J. Ford,
Lect. Notes Phys. {\bf 93}, 334 (1979)

\bibitem{Nakamura}
K. Nakamura and T. Harayama, {\it Quantum Chaos and Quantum Dots}, Oxford University Press, New York, 2004.

\bibitem{Bushong} N. Bushong, Y.V. Pershin, and M. Di Ventra, Phys. Rev. Lett. {\bf 99}, 226802 (2007).

\bibitem{book} M. Di Ventra, {\it Electrical transport in nanoscale systems}, (Cambridge University Press, Cambridge, 2008).

\bibitem{Awschalom}
T. Dietl, D.D. Awschalom, M. Kaminska, and H. Ohno, {\it Spintronics}, (Elsevier, Amsterdam), 2008.

\bibitem{Dyakonov}
{\it Spin physics in semiconductors}, ed. by M.I. Dyakonov, Springer-Verlag Berlin Heidelberg, 2008.

\bibitem{Zutic}
I. Z\v{u}ti\'{c}, J. Fabian, and S. Das Sarma, Rev. Mod. Phys. {\bf 76}, 323 (2004).

\bibitem{Rashba1}
E. I. Rashba and Al. L. Efros, Phys. Rev. Lett. {\bf 91}, 126405 (2003).

\bibitem{Nowack}
K.C. Nowack, F.H.L. Koppens, Yu.V. Nazarov, and L.M.K. Vandersypen,
Science {\bf 318}, 1430 (2007).

\bibitem{Pioro}
M. Pioro-Ladriere, T. Obata, Y. Tokura, Y.-S. Shin, T.Kubo, K. Yoshida, T. Taniyama, and S. Tarucha,
Nature Physics {\bf 4}, 776 {(2008)}.

\bibitem{Golovach06}
V.N. Golovach, M. Borhani, and D. Loss,
Phys. Rev. B {\bf 74}, 165319 (2006).

\bibitem{Levitov03}
L.S. Levitov and E.I. Rashba, Phys. Rev. B {\bf 67}, 115324 (2003).

\bibitem{KS2009}
D. V. Khomitsky and E. Ya. Sherman,  Phys. Rev. B {\bf 79}, 245321 (2009).

\bibitem{KGS2012}
D. V. Khomitsky, L.V. Gulyaev, and E. Ya. Sherman,  Phys. Rev. B {\bf 85}, 125312 (2012).

\bibitem{Chototlishvili}
L. Chotorlishvili, Z. Toklikishvili, A. Komnik, and J. Berakdar, Phys. Lett. A {\bf 377}, 69 (2012).

\bibitem{BerggrenSadreev}
K.-F. Berggren, A.F. Sadreev, and A.A. Starikov, Phys. Rev. E {\bf 66}, 016218 (2002).

\bibitem{BerggrenSadreev2}
K.-F. Berggren, D.N. Maksimov, A.F. Sadreev, R. H{\" o}hmann, U. Kuhl, and H.-J. St{\" o}ckmann,
Phys. Rev. E {\bf 77}, 066209 (2008).

\bibitem{Berggren}
K.-F. Berggren and T. Ouchterlony, Found. Phys. {\bf 31}, 233 (2001).
For a detailed analysis of semiclassical eigenstates in Rashba billiards, see
A. Csord{\'a}s, J. Cserti, A P{\'a}lyi, and U. Z{\"u}licke, Eur. Phys. J. B {\bf 54}, 189 (2006).
For the analysis of spin-orbit coupling effects on level statistics and electromagnetic response
of metal nanoparticles see, for example,
L.P. Gor'kov and G.M. {\'E}liashberg, Sov. Phys. JETP {\bf 21}, 940 (1965).

\bibitem{DM2002}
V.Ya. Demikhovskii, F.M. Izrailev, and A.I. Malyshev,
Phys. Rev. Lett. {\bf 88}, 154101 (2002); Phys. Rev. E {\bf 66}, 036211 (2002);
A.I. Malyshev and L.A. Chizhova, J. Exp. Theor. Phys. {\bf 110}, 837 (2010).

\bibitem{Ritchie} R. Crook, C.G. Smith, A.C. Graham, I. Farrer, H.E. Beere, and D.A. Ritchie,
                  Phys. Rev. Lett. {\bf 91}, 246803 (2003).

\bibitem{BS2004}
E.N. Bulgakov and A.F. Sadreev, Phys. Rev. E {\bf 70}, 056211 (2004).

\bibitem{Ngo2011}
A.T. Ngo, E.H. Kim, and S.E. Ulloa, Phys. Rev. B {\bf 84}, 155457 (2011).

\bibitem{Urbina2013}
J.D. Urbina, M. Wimmer, D. Bauernfeind, D. Espitia, {\. I} Adagideli,
and K. Richter, Phys. Rev. E {\bf 87}, 042115 (2013).

\bibitem{restunn}
A.F. Sadreev, E.N. Bulgakov, and I. Rotter, Phys. Rev. B {\bf 73}, 235342 (2006);
E.N. Bulgakov and I. Rotter, Phys. Rev. E {\bf 73}, 066222 (2006);
{\. I}. Adagideli, Ph. Jackoud, M. Scheid, M. Duckheim, D. Loss, and K. Richter,
Phys. Rev. Lett. {\bf 105}, 246807 (2010);
D. Waltner, J. Kuipers, Ph. Jacqoud, and K. Richter, Phys. Rev. B {\bf 85}, 024302 (2012).


\end{thebibliography}
\end{document}